\documentclass[twoside]{article}
\usepackage[a4paper]{geometry}
\usepackage[latin1]{inputenc}
\usepackage[T1]{fontenc} 
\usepackage{RR}
\usepackage{amsmath}
\usepackage{bbm}
\usepackage{RR}
\usepackage{amssymb}
\usepackage{graphicx}
\usepackage{listings}
\usepackage{multirow}
\usepackage{algorithm}
\usepackage{algorithmic}
\usepackage{url}
\usepackage{hyperref}
\hypersetup{linktocpage}
\usepackage{verbatim}
\usepackage{tikz}
\usetikzlibrary{arrows}
\usepackage{longtable}
\usepackage{caption}
\usepackage{color}
\usepackage{xcolor}
\usepackage{colortbl}
\usepackage{booktabs}
\usepackage{parcolumns}
\usepackage{amsfonts}

\definecolor{lincolngreen}{rgb}{0.11, 0.35, 0.02}
\definecolor{limegreen}{rgb}{0.2, 0.8, 0.2}
\definecolor{napiergreen}{rgb}{0.16, 0.5, 0.0}
\definecolor{beaver}{rgb}{0.62, 0.51, 0.44}
\definecolor{battleshipgrey}{rgb}{0.52, 0.52, 0.51}
\definecolor{bole}{rgb}{0.47, 0.27, 0.23}

\newcommand{\ra}[1]{\renewcommand{\arraystretch}{#1}}
\hypersetup{
  linktocpage,
  colorlinks=true,
  citecolor=blue,
  filecolor=black,
  linkcolor=blue,
  urlcolor=blue
}

\newcommand{\figref}{Fig. \ref}
\newcommand{\lstref}{Listing \ref}

\newtheorem{theorem}{Theorem}[section]

\newtheorem{proposition}[theorem]{Proposition}

\newenvironment{definition}[1][Definition]{\begin{trivlist}
\item[\hskip \labelsep {\bfseries #1}]}{\end{trivlist}}

\newcommand{\qed}{\nobreak \ifvmode \relax \else
      \ifdim\lastskip<1.5em \hskip-\lastskip
      \hskip1.5em plus0em minus0.5em \fi \nobreak
      \vrule height0.75em width0.5em depth0.25em\fi}

\RRNo{8762}
\RRdate{July 2015}
\RRauthor{
Van Chan Ngo
   \and 
Axel Legay
}
\authorhead{Ngo \& Legay}
\RRtitle{Dependability Analysis of Control Systems using SystemC and Statistical Model Checking}
\RRetitle{Dependability Analysis of Control Systems using SystemC and Statistical Model Checking}
\titlehead{Dependability Analysis of Control Systems using SystemC and Statistical Model Checking}
\RRresume{
Petri nets stochastiques sont couramment utilis\'es pour la mod\'elisation de syst\`emes distribu\'es afin d'\'etudier leur performance et fiabilit\'e. Cet article propose une r\'ealisation de Petri nets stochastiques en SystemC pour la mod\'elisation de grands syst\`emes de contr\^ole embarqu\'es. Puis statistical model checking est utilis\'e pour analyser la fiabilit\'e du mod\`ele construit. Notre cadre de v\'erification permet aux utilisateurs d'exprimer une large gamme de propri\'et\'es utiles \`a v\'erifier qui est illustr\'ee par une case-study.
}
\RRabstract{
Stochastic Petri nets are commonly used for modeling distributed systems in order to study their performance and dependability. This paper proposes a realization of stochastic Petri nets in SystemC for modeling large embedded control systems. Then statistical model checking is used to analyze the dependability of the constructed model. Our verification framework allows users to express a wide range of useful properties to be verified which is illustrated through a case study.
}
\RRmotcle{SystemC, Statistical Model Checking, Formal Verification, Dependability Analysis, Petri Nets}
\RRkeyword{SystemC, Statistical Model Checking, Formal Verification, Dependability Analysis, Petri Nets}
\RRprojet{ESTASYS}
\RCRennes 

\begin{document}
\makeRR   

\tableofcontents
\newpage

\section{Introduction} 
\label{sec:introduction}
Computer-based control systems are increasingly used now in a wide range of industrial and military domains such as manufacturing, transport, energy and defense. In many cases, they are safety-critical systems (e.g., control systems for air-traffic, power plants, medical devices). Hence, it is necessary to quantify the probability or rate of all safety-related faults: How likely the system is available to meet a demand for service? What is the probability that the system repairs itself after a failure (e.g., the system conforms to existent and prominent standards such as the Safety Integrity Levels)? A general approach for performing such dependability analysis consists in constructing and analyzing a state-based model of the system~\cite{mcb84,gaa87}. One of the main approaches, \textit{Probabilistic Model Checking} (PMC), is an automatic technique for checking whether or not probabilistic models satisfy certain specifications, which is widely used to verify timed and probabilistic systems~\cite{hkn06,khh09}. One of the main challenges is the complexity of the algorithms in terms of execution time and memory space. Indeed, such algorithms suffer from the state space explosion problem, that is, the size of the state space tends to grow exponentially faster than the size of the system. As a result, the analysis of large systems is infeasible.

An alternative way to evaluate these systems is \textit{Statistical Model Checking} (SMC), a simulation-based approach. Simulation-based approaches do not construct all the reachable states of the system-under-verification (SUV), thus they require far less execution time and memory space than numerical approaches. They have shown the advantages over other methods such as PMC on several case studies~\cite{jcj09,hwz08}.

In this work, we construct a SMC-based verification framework to analyze dependability of large industrial embedded control systems. Stochastic high-level Petri nets (SHLPNs) are traditionally used for modeling distributed control systems in order to study their performance and dependability. Therefore, we propose an approach to model the system dependability by realizing SHLPNs in SystemC.

We then analyze the constructed model in SystemC with Plasma Lab~\cite{bcl13}, a statistical model checker for stochastic processes, in which the properties to be verified are expressed in \textit{Bounded Linear Temporal Logic} (BLTL). The implementation contains two main components: a \textit{monitor generator} that instruments the SystemC model to generate the set of execution traces, and a \textit{checker} that verifies the satisfaction of the properties based on the given set of execution traces. The monitor generation relies on the techniques proposed by Tabakov et al.~\cite{tva10} that provide a rich set of primitives for exposing different parts of the model state during a SystemC simulation.

The remainder of this paper is organized as follows: the next section introduces the SystemC modeling language and reviews the main features of SMC. We consider the execution traces of a SystemC model and the implementation of our verification framework in Section \ref{sec:smcsystemc} and Section \ref{sec:implementation}. An approach to model the dependability of computer-based control systems is proposed in Section \ref{sec:dependability}. Section \ref{sec:casestudy} illustrates the modeling approach and the verification procedure by a case study. The paper terminates with some related work, a conclusion and an outlook to some directions for future research.

\section{Background}
\label{sec:background}
This section introduces the SystemC modeling language and reviews the main features of statistical model checking for stochastic processes.
\subsection{The SystemC Language}
SystemC\footnote{IEEE Standard 1666-2005} is a C++ library~\cite{glm02} providing primitives for modeling hardware and software systems at the level of transactions. Every SystemC model can be compiled with a standard C++ compiler to produce an executable program called executable specification. This specification is used to simulate the system behavior with the provided event-driven simulator.
\subsubsection{Language Features}
A SystemC model is hierarchical composition of modules (\textit{sc\_module}). Modules are building blocks of SystemC design, they are like modules in Verilog~\cite{tmo08}, classes in C++. A module consists of an interface for communicating with other modules and a set of processes running concurrently to describe the functionality of the module. An interface contains ports (\textit{sc\_port}), they are similar to the hardware pins. Modules are interconnected using either primitive channels (i.e., the signals, \textit{sc\_signal}) or hierarchical channels via their ports. Channels are data containers that generate events in the simulation kernel whenever the contained data changes. 

Processes are not hierarchical, so no process can call another process directly. A process is either a thread or a method. A thread process (\textit{sc\_thread}) can suspend its execution  by calling the library statement \textit{wait} or any of its variants. When the execution is resumed, it will continue from that point.  Threads run only once during the execution of the program an are not expected to terminate. On the other hand, a method process (\textit{sc\_method}) cannot suspend its execution by calling \textit{wait} and is expected to terminate. Thus, it only returns the control to the kernel when reaching the end of its body.

An event is an instance of the SystemC event class (\textit{sc\_event}) whose occurrence triggers or resumes the execution of a process. All processes which are suspended by waiting for an event are resumed when this event occurs, we say that the event is notified. A module's process can be sensitive to a list of events. For example, a process may suspend itself and wait for a value change of a specific signal. Then, only this event occurrence can resume the execution of the process. In general, a process can wait for an event, a combination of events, or for an amount time to be resumed.
\subsubsection{SystemC Simulation}
In SystemC, integer values are used as discrete time model. The smallest quantum of time that can be represented is called \textit{time resolution} meaning that any time value smaller than the time resolution will be rounded off. The available time resolutions are femtosecond, picosecond, nanosecond, microsecond, millisecond, and second. SystemC provides functions to set time resolution and declare a time object, for example, the following statements set the time resolution to 10 picosecond and create a time object \textit{t1} representing 20 picoseconds.
\begin{lstlisting}[mathescape]
sc_set_time_resolution(10, SC_PS); 
sc_time t1(20, SC_PS); // SC_PS : picosecond
\end{lstlisting}

The SystemC simulator is an event-driven simulation~\cite{sysc,mrh01}. It establishes a hierarchical network of finite number of parallel communicating processes which under the supervision of the distinguished simulation kernel process. Only one process is dispatched by the scheduler to run at a time point, and the scheduler is non-preemptive, that is, the running process returns control to the kernel only when it finishes executing or explicitly suspends itself by calling \textit{wait}. Like hardware modeling languages, the SystemC scheduler supports the notion of delta-cycles~\cite{lsu93}. A delta-cycle lasts for an infinitesimal amount of time and is used to impose a partial order of simultaneous actions which interprets zero-delay semantics. Thus, the simulation time is not advanced when the scheduler processes a delta-cycle. During a delta-cycle, the scheduler executes actions in two phases: the \textit{evaluate} and the \textit{update} phases. 

The simulation semantics of the SystemC scheduler is presented as follows: (1) \textit{Initialize}. During the initialization, each process is executed once unless it is turned off by calling \textit{dont\_initialize()}, or until a synchronization point (i.e., a \textit{wait}) is reached. The order in which these processes are executed is unspecified; (2) \textit{Evaluate}. The kernel starts a delta-cycle and run all processes that are ready to run one at a time. In this same phase a process can be made ready to run by an event notification; (3) \textit{Update}. Execute any pending calls to \textit{update()} resulting from calls to \textit{request\_update()} in the evaluate phase. Note that a primitive channel uses \textit{request\_update()} to have the kernel call its \textit{update()} function after the execution of processes; (4) \textit{Delta-cycle notification}. The kernel enters the delta notification phase where notified events trigger their dependent processes. Note that immediate notifications may make new processes runable during step (2). If so the kernel loops back to step (2) and starts another evaluation phase and a new delta-cycle. It does not advance simulation time; (5) \textit{Simulation-cycle notification}. If there are no more runnable processes, the kernel advances simulation time to the earliest pending timed notification. All processes sensitive to this event are triggered and the kernel loops back to step (2) and starts a new delta-cycle. This process is finished when all processes have terminated or the specified simulation time is passed. The simulation semantics can be represented by the pseudo code in \lstref{fig:simulationsemantics}.
\begin{lstlisting}[caption=Simulation semantics of SystemC,label=fig:simulationsemantics,mathescape]
PC // All primitive channels
P // All processes
R = $\emptyset$ // Set of runnable processes
D = $\emptyset$ // Set of pending delta notifications
U = $\emptyset$ // Set of update requests
T = $\emptyset$ // Set of pending timed notifications
// Start elaboration: collect all update requests in U
for all chan $\in$ U do
  run chan.update()
end for
for all p $\in$ P do
  if p is initialized and p is not clocked thread then
    R = R $\cup$ p // Make p runnable
  end if
end for
for all p $\in$ P do
  if p is triggered by an event in D then 
    R = R $\cup$ p
  end if
end for // End of initialization phase

repeat
  while R $\neq$ $\emptyset$ do // New delta-cycle begins
    for all r $\in$ R do // Evaluation phase
      R = R $\setminus$ r
      run r until it calls wait or returns
    end for
    for all chan $\in$ U do // Update phase
      run chan.update()
    end for
    for all p $\in$ P dp // Delta notification phase
      if p is triggered by an event in D then
        R = R $\cup$ p // Make p runnable
      end if
    end for // End of delta-cycle
  end while
   
  if T $\neq$ $\emptyset$ then
    Advance the simulation clock to the earliest timed delay t
    T = T $\setminus$ t      
    for all p $\in$ P do // Timed notification phase
      if t triggers p then
        R = R $\cup$ p // Make p runnable
      end if
    end for
  end if
until end of simulation
\end{lstlisting}
\subsection{Statistical Model Checking}
Let $\mathcal{M}$ be a model of a stochastic process and $\varphi$ be a property expressed as a BLTL formula. 
BLTL is a temporal logic with bounded temporal operators, ensuring that the satisfaction of a formula by a trace can be decided in a finite number of steps.
The statistical probabilistic model checking problem consists in answering the following questions. 
\begin{itemize}
\item \textit{Qualitative}. Is the probability that $\mathcal{M}$ satisfies $\varphi$ greater or equal to a threshold $\theta$ with a specific level of statistical confidence? 
\item \textit{Quantitative}. What is the probability that $\mathcal{M}$ satisfies $\varphi$ with a specific level of statistical confidence? 
\end{itemize}
They are denoted by $\mathcal{M} \models Pr(\varphi)$ and $\mathcal{M} \models Pr_{\geq \theta}(\varphi)$, respectively. 

The key idea of SMC~\cite{ldb10} is to get, through simulation, a large amount of independent execution traces and count the number of traces that satisfy $\varphi$. The ratio of this number over the total number of execution traces provides the probability that the property holds. 
Then statistical results associate a level of confidence to this probability, depending on the number of execution traces. 
Many statistical methods including sequential hypothesis testing, Monte Carlo method, or 2-sided Chernoff bound are implemented in a set of existing tools~\cite{yhl05,bcl13}, that have shown their advantages over other methods such as PMC on several case studies.

Although SMC can only provide approximate results with a user-specified level of statistical confidence (as opposed to the exact results provided by standard probabilistic model checking method), it is compensated for by its better scalability and resource consumption. Since the models to be analyzed are often approximately known, an approximate result in the analysis of desired properties within specific bounds is quite acceptable. SMC has recently been applied in a wide range of research areas including software engineering (e.g., verification of critical embedded systems)~\cite{hwz08}, system biology, or medical area~\cite{jcj09}.

\section{SMC for SystemC Models}
\label{sec:smcsystemc}
This section illustrates the use of SMC for verifying a SystemC program exhibiting timed and probabilistic characteristics by showing that the operational semantics of the program is viewed as a stochastic process. The implementation of our SMC-based verification framework is considered as well. 
\subsection{SystemC Model State}
Temporal logic formulas are interpreted over execution traces and traditionally a trace has been defined as a sequence of states in the execution of the model. Therefore before we can define an execution trace we need a precise definition of the state of a SystemC simulation. We are inspired by the definition of system state in~\cite{tva10}, which consists of the state of the simulation kernel and the state of the SystemC model. We consider the external libraries as black boxes, meaning that their states are not exposed. 

The state of the kernel contains the information about the current phase of the simulation and the SystemC events notified during the execution of the model. We denote the finite set of variables whose value domain represent the information about the current phase of the kernel by $V_{ker}$, and the finite set of variables whose value domain represent the event notification by $V_{eve}$. 

The state of the SystemC model is the full state of the C++ code of all processes in the model, which includes the values of the variables, the location of the program counter, the call stack, and the status of the processes. We use $V_{var}, V_{loc}, V_{sta}$, and $V_{proc}$ to denote the finite sets of variables whose value domains represent the values of the variables, the location of the program counter, the call stack and the status of the processes, respectively. Let $V = \{v_0,...,v_{n-1}\} = \bigcup_k V_k$ with $k \in \{ker, eve, var, loc, sta, proc\}$, be a finite set of variables that takes values in a domain $\mathbb{D}_X$. A value in $\mathbb{D}_X$ represents a state of a SystemC simulation. 

We consider here some examples of variables in $V$ that represent the state of the simulation kernel and the SystemC model. A system state can consists of the information about all locations before the execution of all statements that contain the devision operator ``/'' followed by zero or more spaces and the variable ``\textit{a}'' in a SystemC model (e.g., the statement \textit{y = (x + 1) / a}). Let \textit{Producer} and \textit{Consumer} be two modules of a SystemC model. Assume that \textit{Producer} has a function \textit{send()} and \textit{Consumer} has a function \textit{receive()}. Two variables \textit{send\_start} and \textit{send\_done} can be defined as Boolean variables in $V$ that hold the value \textit{true} immediately before and after a call of the function \textit{send()}, respectively to express the status of the call stack. Similarly, the variable \textit{rcv} holds the value \textit{true} immediately after a call of the function \textit{receive()}. Assume again that \textit{Producer} has an event named \textit{write\_event}, to observe whenever this event is notified, we can use a variable \textit{we\_notified} that holds the value \textit{true} immediately when the event is notified. We will consider how these variables can be defined in our implementation of the verification framework in the next section.

We have discussed so far the state of a SystemC model execution. It remains to discuss how the semantics of the temporal operators is interpreted over the states in the execution of the model. That means how the states are sampled. The following definition gives the concept of \textit{temporal resolution}, in which the states are evaluated only in instances in which the temporal resolution holds. It allows the user to set granularity of time. 
\begin{definition}[Temporal resolution]
A temporal resolution $\mathcal{T}_r$ is a finite set of Boolean expressions defined over $V$ which specifies when the set of variables $V$ is evaluated.
\end{definition}
Temporal resolution can be used to define a more fine-grained model of time than a coarse-grained one provided by a cycle-based simulation. We call the expressions in $\mathcal{T}_r$ \textit{temporal events}. Whenever a temporal event is satisfied or the temporal event occurs, $V$ is sampled. For example, assume that we want the set of variables to be sampled whenever at the end of simulation-cycle notification or immediately after the event \textit{write\_event} is notified during a run of the model. Hence, we can define a temporal resolution as the following set $\mathcal{T}_r =$ $\{e_{dc}, we\_notified\}$, where $e_{dc}$ and \textit{we\_notified} are Boolean expressions that have the value \textit{true} whenever the kernel phase is at the end of the simulation-cycle notification and the event \textit{write\_event} notified, respectively.

We denote the set of occurrences of temporal events from $\mathcal{T}_r$ along an execution of a SystemC model by $\mathcal{T}^{s}_r$, called a \textit{temporal resolution set}. The value of a variable $v \in V$ at an event occurrence $e_c \in \mathcal{T}^{s}_{r}$ is defined by a mapping $\xi^{v}_{val}: \mathcal{T}^{s}_{r} \rightarrow \mathbb{D}_{v}$. Hence, the state of the SystemC model at $e_c$ is defined by a tuple $(\xi^{v_0}_{val},...,\xi^{v_{n-1}}_{val})$.

A mapping $\xi_t: \mathcal{T}^{s}_{r} \rightarrow \mathcal{T}$ is called a \textit{time event} that identifies the simulation time at each occurrence of an event from the temporal resolution. Hence, the set of time points which correspond to a temporal resolution set $\mathcal{T}^{s}_{r} = \{e_{c_0},...,e_{c_{N-1}}\}, N \in \mathbb{N}$ is given as follows.
\begin{definition}[Time tag]
Given a temporal resolution set $\mathcal{T}^{s}_{r}$, the \textit{time tag} $\mathcal{T}$ corresponding to $\mathcal{T}^{s}_{r}$ is a finite or infinite set of non-negative reals $\{t_0,t_1,...,t_{N-1}\}$, where $t_{i+1} - t_i = \delta t_i \in \mathbb{R}_{\geq 0}, t_i = \xi_t(e_{c_i})$.
\end{definition}
\subsection{Model and Execution Trace}
A SystemC model can be viewed as a hierarchical network of parallel communicating processes. Hence the execution of a SystemC model is an alternation of the control between the model's processes, the external libraries and the kernel process. The execution of the processes is supervised by the kernel process to concurrently update new values for the signals and variables w.r.t the cycle-based simulation. For example, given a set of runnable processes in a simulation-cycle, the kernel chooses one of them to execute first in a non-deterministic manner as described in the prior section.

Let $V$ be the set of variables whose values represent the state of a SystemC model simulation. Our way to mix stochastic and non-deterministic characteristics consisting of assuming that, at any moment of time, the values variables in $P \subseteq V$ are determined by a given probability distribution (i.e., from the probability distributions used in the model). The values of variables in $V \setminus P$ are chosen in the non-deterministic manner of the simulation scheduler. At any given moment of time, it is allowed that the choice of the variables in $V \setminus P$ might influence the distribution on the next values of variables in $P$.

Given a temporal resolution $\mathcal{T}_r$ and its corresponding temporal resolution set along an execution of the model $\mathcal{T}^{s}_{r} = \{e_{c_0},...,e_{c_{N-1}}\}, N \in \mathbb{N}$, the evaluation of $V$ at the event occurrence $e_{c_i}$ is defined by the tuple $(\xi^{v_0}_{val},...,\xi^{v_{n-1}}_{val})$, or a state of the model, denoted by $V(e_{c_i}) = (V(e_{c_i})(v_0),V(e_{c_i})(v_1),...,V(e_{c_i})(v_{n-1}))$, where $V(e_{c_i})(v_k) = \xi^{v_k}_{val}(e_{c_i})$ with $k = 0,...,n-1$ is the value of the variable $v_k$ at $e_{c_i}$. We denote the set of all possible evaluations by $V_{\mathcal{T}^{s}_{r}} \subseteq \mathbb{D}_V$, called the \textit{state space} of the random variables in $V$. State changes are observed only at the moments of event occurrences. Hence, the operational semantics of a SystemC model is represented by a \textit{stochastic process} $\{(V(e_{c_i}),\xi_t(e_{c_i})), e_{c_i} \in \mathcal{T}^{s}_{r}\}_{i \in \mathbb{N}}$, taking values in $V_{\mathcal{T}^{s}_{r}} \times \mathbb{R}_{\geq 0}$ and indexed by the parameter $e_{c_i}$, which are event occurrences in the temporal resolution set $\mathcal{T}^{s}_{r}$. An execution trace is a realization of the stochastic process is given as follows.
\begin{definition}[Execution trace]
An execution trace of a SystemC model corresponding to a temporal resolution set $\mathcal{T}^{s}_{r} = \{e_{c_0},...,e_{c_{N-1}}\}, N \in \mathbb{N}$ is a sequence of states and event occurrence times, denoted by $\omega = (s_0,t_0)(s_1,t_1)...(s_{N-1},t_{N-1})$, such that for each $i \in 0,...,N-1$, $s_i = V(e_{c_i})$ and $t_i = \xi_t(e_{c_i})$. 
\end{definition}
$N$ is the length of the execution, also denoted by $|\omega|$. We denote the prefix of $\omega$ by $\omega_k = (s_0,t_0),(s_1,t_1)...(s_k,t_k)$, and the suffix by $\omega^k = (s_k,t_k)(s_{k+1},t_{k+1})...(s_{N-1},t_{N-1})$. 

Let $V' \subseteq V$, the \textit{projection} of $\omega$ on $V'$, denoted by $\omega \downarrow_{V'}$, is an execution trace such that $|\omega \downarrow_{V'}| = |\omega|$ and $\forall v \in V'$, $\forall e_c \in \mathcal{T}^{s}_{r}$, $V'(e_c)(v) = V(e_c)(v)$.
\subsection{Expressing Properties}
\label{sec:bltl}
We recall the syntax and semantics of BLTL~\cite{sva05}, an extension of Linear Temporal Logic (LTL) with time bounds on temporal operators. A BLTL formula $\varphi$ is defined over a set of atomic propositions $AP$ as in LTL. A BLTL formula is defined by the grammar $\varphi ::= true | false | p \in AP | \varphi_1 \wedge \varphi_2 | \neg \varphi | \varphi_1 \; U_{\leq T} \; \varphi_2$, where the time bound $T$ is an amount of time or a number of states in the execution trace. The temporal modalities $F$ (the ``eventually'', sometimes in the future) and $G$ (the ``always'', from now on forever) can be derived from the ``until'' $U$ as follows.
\begin{displaymath}
F_{\leq T} \; \varphi = true \; U_{\leq T} \; \varphi \text{ and } G_{\leq T} \; \varphi = \neg F_{\leq T} \; \neg \varphi
\end{displaymath}
The semantics of BLTL is defined w.r.t execution traces of the model $\mathcal{M}$. Let $\omega$ be an execution trace of $\mathcal{M}$, we denote by $\omega \models \varphi$ that fact that $\omega$ satisfies the BLTL formula $\varphi$. 
\begin{itemize}
\item $\omega^k \models true$ and $\omega^k \not \models false$
\item $\omega^{k} \models p, p \in AP$ iff $p \in L(s_k)$, where $L(s_k)$ is the set of atomic propositions which are $true$ in state $s_k$ 
\item $\omega^{k} \models \varphi_1 \wedge \varphi_2$ iff $\omega^k \models \varphi_1$ and $\omega^k \models \varphi_2$
\item $\omega^k \models \neg \varphi$ iff $\omega^k \not \models \varphi$
\item $\omega^k \models \varphi_1 \; U_{\leq T} \; \varphi_2$ iff there exists an integer $i$ such that $\omega^{k+i} \models \varphi_2$, $\Sigma_{0 < j \leq i}(t_{k+j} - t_{k+j-1}) \leq T$, and for each $0 \leq j < i, \omega^{k+j} \models \varphi_1$
\end{itemize}
In our framework, the set of atomic proposition $AP$ consists of the predicates defined over the set of variables $V$. Using these predicates, users can define temporal properties related to the state of the kernel and the state of the SystemC model. Recall that \textit{Producer} and \textit{Consumer} are two SystemC modules that have two functions \textit{send()} and \textit{receive()}, respectively. We consider the following variables $send\_start$, $send\_done$ and $rcv \in V$. They expose the state of the SystemC model as described in the section above. Assume that we want to express the property \textit{``over a period of $T_1$ time units, \textit{send()} remains blocked until \textit{receive()} has returned within $T_2$ time units''}. This property can be specified with the ``until'' operator that is given as follows.
\begin{displaymath}
G_{\leq T_1}(send\_start \rightarrow(\neg send\_done \; U_{\leq T_2} \; rcv))
\end{displaymath}
\section{Implementation}
\label{sec:implementation}
\begin{figure}[ht]
\begin{center}
\includegraphics[width=0.85\textwidth]{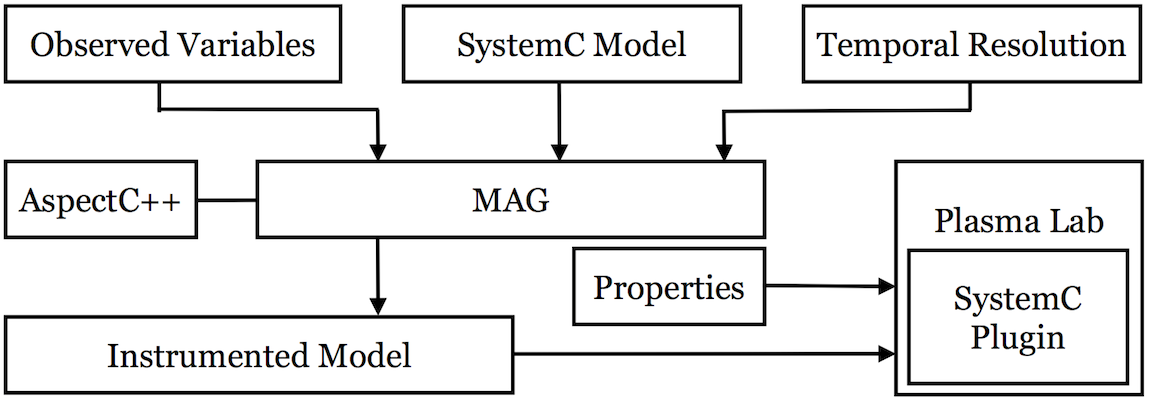}
\caption{The framework's flow}
\label{fig:architecture}
\end{center}
\end{figure}

We have implemented a SMC-based verification framework \cite{nlq14} which is used to analyze the case study in Section \ref{sec:casestudy}. Our implementation contains two main components: a \textit{monitor and aspect-advice generator} (MAG) and a \textit{statistical model checker} (SystemC Plugin). The flow of our framework is depicted in \figref{fig:architecture}.
\subsection{Monitor and Aspect-Advice Generator}
In principle, the full state can be observed during the simulation of the model. In practice, however, users define a set of variables of interest, called \textit{observed variables}, and only these variables appear in the states of an execution trace. Given a SystemC model, an observed variable is a variable of primitive type (e.g, usual scalar or enumerated type in C/C++). The value of this variable represents a part of the full state of the model. We use $V_{obs} \subseteq V$ to denote the set of observed variables. Then, the observed execution traces of the model are the projections of the execution traces on $V_{obs}$, meaning that for every execution trace $\omega$, the corresponding observed execution trace is $\omega \downarrow_{V_{obs}}$. In the following, when we mention about execution traces, we mean observed execution traces.

The implementation of MAG allows users to define a set of observed variables that is used with a temporal resolution to generate a monitor based on the techniques in~\cite{tva10} in order to make an instrumented SystemC model. The instrumented model will produce a set of execution traces of the model. The generated monitor evaluates the set of observed variables at every time point in which an event of the temporal resolution occurs during the SystemC model simulation. The generator also generates an aspect-advice file that is used by AspectC++~\cite{gss01} to automatically instrument the SystemC model.
\subsection{SystemC Plasma Lab Plugin}
Our statistical model checker is implemented as a plugin of Plasma Lab~\cite{bcl13} which establishes an interface between Plasma Lab and the instrumented model being executed by the simulator. In the current version, the communication is done via the standard input and output. The plugin requests new states until the satisfaction of the formula to be verified can be decided, which terminates because the temporal operators are bounded. Similarly, depending on the hypothesis testing algorithms (e.g., sequential hypothesis testing, Monte Carlo simulation, or 2-sided Chernoff bound), the plugin will request new traces from the instrumented model.
\subsection{Running Verification}
Running the verification framework consists of two steps as follows. First, users define a set of observed variables and a temporal resolution in a configuration file, as well as other necessary information. From that information, the generator generates the monitors and aspect-advices that are used by AspectC$^{++}$ to produce the instrumented SystemC model. In addition, the generator can automatically generate a Plasma Lab project file according to the desired properties. The instrumented model and the generated monitors are compiled together and linked with the SystemC simulation kernel into an executable model in order to make a set of execution traces of the system. In the second step, the plugin of Plasma Lab is used to verify the desired properties. The satisfaction checking of the properties is brought out based on the set of execution traces by executing the instrumented SystemC model and can be done by several hypothesis testing algorithms provided by Plasma Lab. The full implementation of our verification framework including the monitor and aspect-advice generator and the checker can be downloaded on the website of Plasma Lab\footnote{MAG manual: \url{https://project.inria.fr/plasma-lab/documentation/tutorial/mag_manual/}}.

\section{Modeling Dependability in SystemC}
\label{sec:dependability}
SHLPNs are high-level Petri nets (HLPNs)~\cite{mur89,jro91,mol82}, in which each transition execution has a duration described by an exponential distribution. 
They are commonly used for modeling distributed systems 
in order to study their performance and dependability~\cite{gaa87,mol82,mcb84}. In this section, we propose an approach for realizing SHLPNs in SystemC such that the  semantics is preserved.
\subsection{Stochastic High-Level Petri Nets}
High-level Petri nets provide a compact representation of complex systems. There are many different types of HLPNs that have been proposed in literature such as predicate transition nets~\cite{gen79}, coloured Petri nets~\cite{jen81} and relation nets~\cite{rei83}. However, in~\cite{jen83,rei83} it is proved that one can translate the HLPN of a system in one type into any other type. Due to the intuition and the modeling elegance we consider the predicate transition nets in which the tokens are \textit{coloured} or \textit{typed tokens}~\cite{jro91}. Each place is annotated with a data type. Each place is associated with a \textit{place capacity} $K$ which bounds for the number of 
tokens that the place can contain. The arcs are labeled by tuples of token variables. The zero-tuple indicates an ordinary place, meaning a no-argument token. The transitions are annotated with \textit{guard formulas} defined over variables labelling the adjacent arcs and a set of \textit{variable assignments}. A transition is \emph{enabled}, that is may execute or \emph{fire}, whenever 1) all input places carry enough tokens, 2) there is an assignment of tokens to token variables that satisfies the guard formula of the transition,  and 3) each output place capacity is sufficient to store the tokens produced by the transition. By firing, a transition removes and adds tokens from/to places according to the expressions labeling the arcs. SHLPNs are high-level Petri nets in which each execution of a transition lasts for a given amount of time, called \textit{firing time}. Firing times are specified by an exponential distribution associated to each transition. 
Due to the memoryless property of the exponential distribution, this class of SHLPNs is isomorphic to continuous-time Markov chains (CTMCs) as shown in~\cite{mol82}.
\begin{figure}[ht]
\centering
\includegraphics[width=0.80\textwidth]{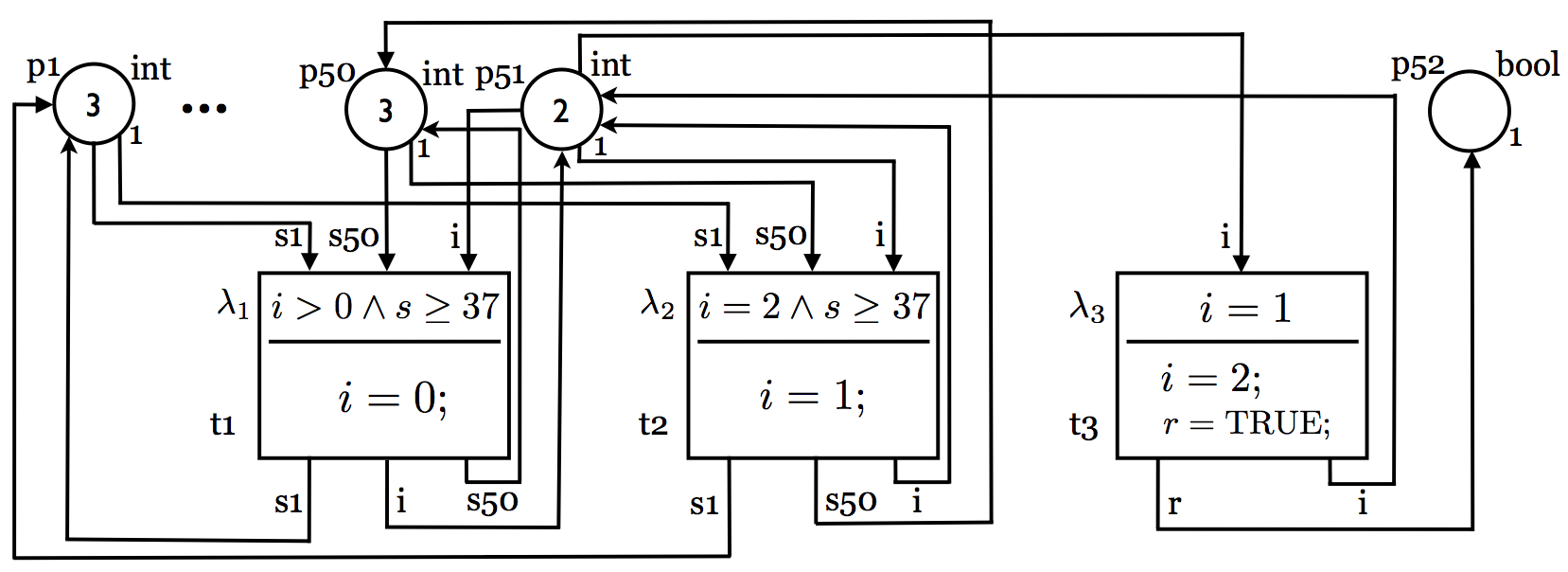}
\caption{Stochastic high-level Petri net example}
\label{fig:shlpn}
\end{figure}
We consider the example in Fig. \ref{fig:shlpn}. This net is a part of the SHLPN of our case study models the reliability of the input processor. The example consists of $51$ I/O places with data type $\begin{tt}int\end{tt}$ ($p_1$ to $p_{50}$ represent the number of functional sensors in each group and $p_{51}$ representing the current status of the processor) and output place $p_{52}$ with data type $\begin{tt}bool\end{tt}$ carries a token with the value $\begin{tt}true\end{tt}$ whenever the processor reboots successfully. Their capacities are $1$. Each of transitions $t_1, t_2$ and $t_3$ is annotated with a firing time following the exponential distribution with the rate $\lambda_i, i = 1,...,3$. These rates may be marking-dependent. With the marking depicted in Fig. \ref{fig:shlpn}, $s_1$ to $s_{50}$, and $i$ are assigned values $3$ and $2$ respectively, both transitions $t_1$ and $t_2$ are enabled to fire that means they conflict, where $s = \Sigma^{50}_{k = 1}\sigma_k, \sigma_k = 1 \text{ if } s_k \geq 2, \text{ otherwise } \sigma_k = 0$. Assume that $t_1$ fires, $p_1$ to $p_{50}$ carry the same previous token values, $p_{51}$ caries the value $0$ and $p_{52}$ is empty.
\subsection{Connection between HLPNs and Rule-Based Systems}
A rule-based system (or production based system)~\cite{br85} consists of a \textit{working memory} containing known facts, a \textit{production memory} containing rules, and an \textit{inference engine} which matches the rules with the working memory to infer new facts by applying the selected rule among all the applicable rules and the existing facts. The rules syntactically are of the form \textit{``if conditions then actions''}. The conditions are patterns that are checked for the rule activation, called the rule \textit{antecedent}. If the conditions match with the facts, the actions, called the rule \textit{consequent}, are performed. Fig. \ref{fig:inference} depicts the main structure of a forward chaining inference algorithm.
\begin{figure}[ht]
\centering
\begin{minipage}{.49\textwidth}
\begin{lstlisting}[mathescape]
 void Infer(working_memory,production_memory) {
  rules = Select(working_memory,production_memory);
  while(rules $\not = \emptyset$) {
   rule = SolveConflicts(rules);
   ApplyRule(rule);
   rules  = Select(working_memory,production_memory);
  } //END while
 }
\end{lstlisting}
\end{minipage}
\begin{minipage}{.49\textwidth}
\begin{lstlisting}[firstnumber=1,mathescape]
 rule_set Select(working_memory,production_memory) {
  rules = $\emptyset$;
  for rule $\in$ production_memory {
   if Match(rule,working_memory)
    rules = rules $\cup$ rule;
  } //END for
  return rules;
 }
\end{lstlisting}
\end{minipage}
\caption{A forward chaining inference algorithm}
\label{fig:inference}
\end{figure}
The main function, \texttt{Infer}, executes the rule system until no more rule can be executed.
The function \texttt{Select} identifies a set of rules that matches the facts in the working memory, according to the \texttt{Match} function, which maps the variables appearing into the conditions to some constants from the facts of the working memory. The function \textit{SolveConflicts} selects one of the selected rule to be fired. Finally, the function \textit{ApplyRule} executes the rule actions and updates the working memory.

For every HLPN one can construct a rule-based system, as shown in~\cite{rv90,bbr03}. The transformation consists of the production memory and working memory constructions. For each transition of the net, a rule is added in which the conditions and actions are defined by the guard formula, the set of variable assignments, the input and output places of the transition. For each input place, the facts that bound the tokens to the variables labeling the arc connecting the place to the transitions are added and the condition expressing that the place carries enough copies of proper tokens, as well as the action of removing tokens from the place are defined for the corresponding rule. For each output place, the action of adding tokens to the place is associated, by specifying the place name and the variables on the arc connecting the transition to the place and the condition expressing that the place's capacity is not exceeded by adding the produced tokens is defined. The actions of adding and removing tokens update the facts in the working memory. The initial working memory is constructed based on the initial marking of the net. In other words, a fact represents a multiset of constants. This set is compatible with the type and capacity of its associated place. The following proposition from~\cite{rv90,bbr03} shows that the HLPN and its corresponding transformed rule-based system are semantically equivalent.
\begin{proposition}
\label{prop:pnrbs}
For every high-level Petri net, a rule-based system exists that is semantic equivalent.
\end{proposition}
The proof of Proposition \ref{prop:pnrbs} and the similarities between HLPNs and rule-based systems have been studied in~\cite{rv90,bbr03,mzh88} by showing that the rules applied by the inference algorithm represent the reachability set from the initial marking of the initial Petri net. It follows that the inference algorithms can be use to implement the operational semantics of the stochastic high-level Petri nets. In addition, it is shown in~\cite{rv90,bbr03} that the implementation of HLPNs semantics based on rule-based systems with an improved version of the inference algorithm was described by Forgy~\cite{for82}, is most efficient when the net have large number of places and large number of tokens distributed among the places, as well as when the large number bounded variables labeling the arcs. In the next section, we propose an approach to realize SHLPNs by implementing the inference algorithms in SystemC.
\subsection{Realization of SHLPNs}
We illustrate our realization of the inference algorithm in Fig. \ref{fig:inference} with the example in Fig. \ref{fig:shlpn}. It is implemented by a SystemC module as shown in Fig. \ref{fig:inferencesystemc}, in which the \textit{Infer} function is a thread process in SystemC. The initialization initialize the working memory by setting all places, the capacity and the initial marking in the constructor method (lines $37$ to $40$). And a place is implemented as an instance of a template class (i.e., \textit{place}$<$\textit{int}$>$) that contains the facts. The template class has the methods for getting a token value (\textit{get}), storing a token (\textit{mark}), removing a token (\textit{demark}), getting the capacity (\textit{get\_capacity}), and the current number of tokens (\textit{get\_num}) in a place.

Consider a SHLPN with the current marking $M_j$, transitions become enabled as usual, i.e., if all input places have sufficiently tokens and the guard formulas are satisfied. However, there is a time, which has to elapse, before an enabled transition fires. We denote the set of all enabled transitions $t_1,...,t_n$ in $M_j$ by $E(M_j)$. Since all firing times are independent exponentially distributed, the minimum firing time is also exponentially distributed with the rate $\lambda = \Sigma^n_{i=1}\lambda_i$ and the probability that a given enabled transition, say $t_k$, samples the minimum firing time $Pr(t_k|M_j)= \lambda_k/\Sigma_{i:t_i \in E(M_j)}\lambda_i$. 
\begin{figure}[ht]
\centering
\begin{minipage}{.49\textwidth}
\begin{lstlisting}[mathescape]
 SC_MODULE(Shlpn) {
  SC_HAS_PROCESS(Shlpn);
 public:
  Shlpn(sc_module_name name, gsl_rng *rnd);
  //operation of the net
  void Infer();
 private:
  place<int> p[51];
  place<bool> p52;
  int i, s[50];
  bool r;
  //firing rates
  double r1, r2, r3, sum_r, ft;
  //GSL random generator
  gsl_rng *rnd;
  //enabled transitions
  bool e[3];
  //return enabled transitions
  int Select();
  int SolveConflicts();
  ApplyRule(int rule);
 };
 void Shlpn::Infer() {
  for (int k = 0; k < 3; k++)
   e[k] = false;
  sum_r = ft = 0;
  int rules = Select();
  while(rules > 0) {
   int rule = SolveConflicts();
   ApplyRule(rule);
   for (int k = 0; k < 3; k++)
    e[k] = false;
   sum_r = ft = 0;
   rules = Select();
  } //END while
 }
\end{lstlisting}
\end{minipage}
\begin{minipage}{.49\textwidth}
\begin{lstlisting}[firstnumber=37,mathescape] 
 Shlpn::Shlpn(sc_module_name name, gsl_rng *rnd) {
  //initialization
  SC_THREAD(Infer):
 }
 int Shlpn::SolveConflicts() {
  int rule;
  double pr1, pr2, pr3;
  //probability t1 fires
  pr1 = e[0] ? r1/sum_r : 0;
  //probabilities of t2, t3
  pr2 = e[1] ? r2/sum_r : 0;
  pr3 = e[2] ? r3/sum_r : 0;
  //fired transition
  rule = gsl_discrete({pr1,pr2,pr3},rnd); 
 }
 void Shlpn::ApplyRule(int rule) {
  switch (rule) {
   case 0: //fire t1
    //elapse firing time
    ft = gsl_exp(sum_r,rnd);
    wait(ft,time_unit);
    for (int k = 0; k < 51; k++)
     p[k].demark();
    i = 0;
    for (int k = 0; k < 50; k++)
     p[k].mark(3);
    p[50].mark(i); 
    break;
   case 1: //fire t2
   case 2: //fire t3
   default: 
    break;
  }// END switch
 }
\end{lstlisting}
\end{minipage}
\caption{SystemC code for example in Figure \ref{fig:shlpn}}
\label{fig:inferencesystemc}
\end{figure}

Therefore, the implementations of the functions \textit{SolveConflicts} and \textit{ApplyRule} is done as follows. Given a set of all enabled rules from the function \textit{Select} in Appendix A, \textit{SolveConflicts} (lines $41$ to $51$) determines a rule to be applied from the set of selected rules using a discrete distribution over $Pr(t_k|M_j)$ (lines $45$ to $50$). The firing time before the firing of selected rule in \textit{ApplyRule} (lines $52$ to $70$), is sampled from the exponential distribution with the rate $\lambda$ (lines $56$ to $57$). We employ the implementation of the discrete and exponential distributions from GNU Scientific Library (GSL). The firing time elapsing is simulated by a \textit{wait()} statement with an amount of time equals to the firing time (i.e., measured by the time unit in the simulator).

The implementation of the function \textit{Select} and the place template class are given in Listing \ref{lst:select} and Listing \ref{lst:intplace}.
\begin{lstlisting}[caption=SystemC code for the \textit{Select} function,label=lst:select,mathescape]
 int Shlpn::Select() {
  int s, rules = 0;
  bool check_p = true;
  //check t1 is enabled
  //check all input places have sufficiently tokens
  for(int k = 0; k < 50; k++) {
   if (p[k]->get_num() > 0)
    s[k] = p[k]->get();
   else {
    check_p = false;
    break;
   }
  } //END for
  if (p[50]->get_num() > 0)
    i = p[50]->get();
  else check_p = false;
  //check the guard formula
  if (check_p) {
   for (int k = 0; k < 50; k++)
    if (s[k] >= 2) 
     s = s + 1;
   if (s >= 37 && i > 0) {
    //set t1 is enabled transition
    e[0] = true;
    sum_r = sum_r + r1;
    rules = rules + 1;
   } //END if
  } //END if
  //check t2 is enabled
  //check t3 is enabled
  return rules;
 }
\end{lstlisting}
\begin{lstlisting}[caption=SystemC code for a place,label=lst:intplace,mathescape]
 template <class type> class place {
 public:
   type get(int); //get token at index
   bool mark(type); //store token
   void demark(int); //delete token at index
   int get_capacity(); //place capacity
   int get_num(); //number of tokens
 };
 
 template <class type> type place<type>::get(int index) {
    //code
 }

 template <class type> bool place<type>::mark(type t) {
    //code
 }
\end{lstlisting}

\section{Case Study and Results}
\label{sec:casestudy}
In this section, our SystemC-realization is used to model the dependability of a large embedded control system. We also demonstrate the use of our verification framework to analyze the resulting model. The number of components in our system makes numerical approaches such as PMC unfeasible. 
\subsection{An Embedded Control System}
\begin{figure}[ht]
\centering
\includegraphics[width=0.7\textwidth]{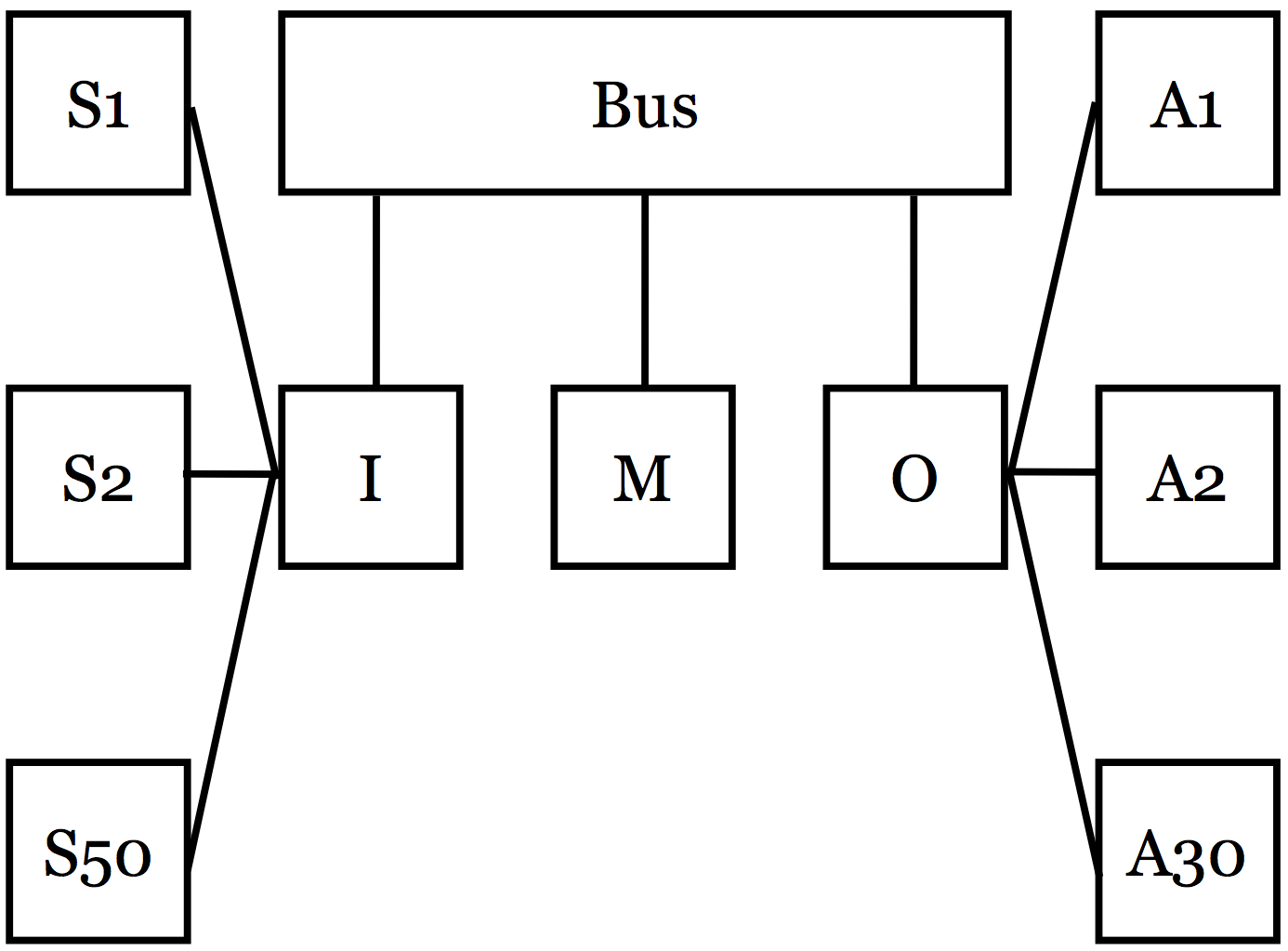}
\caption{A control system}
\label{fig:system}
\end{figure}

The case study is closely based on the one presented in \cite{mct94,knp07} but contains much more components. The system, depicted in Fig. \ref{fig:system}, consists of an input processor ($I$) connected to $50$ groups of $3$ sensors (from $S_1$ to $S_{50}$), an output processor ($O$), connected to $30$ groups of $2$ actuators (from $A_1$ to $A_{30}$), and a main processor ($M$), that communicates with $I$ and $O$ through a bus. At every cycle, the main processor polls results from the input processor that reads and processes data from the sensors. Based on these results, it elaborates commands to be passed to the output processor which controls the actuators. For instance, the input sensors can measure the fluid level, temperature, or pressure, while the commands sent to actuators could be used for controlling valves.

The reliability of the system is affected by the failures of the sensors, actuators, and processors. The probability of bus failure is negligible, hence we do not consider it. The sensors and actuators are used in $37-\text{of}-50$ and $27-\text{of}-30$ modular redundancies, respectively. Hence, if at least $37$ sensor groups are functional (a sensor group is functional if at least $2$ of the $3$ sensors are functional), the system obtains enough information to function properly. Otherwise, the main processor is reported to shut the system down. In the same way, the system requires at least $27$ functional actuator groups to function properly (a actuator group is functional if at least $1$ of the $2$ actuators is functional). Transient and permanent faults can occur in processors $I$ or $O$ and prevent the main processor($M$) to read data from $I$ or send commands to $O$. In that case, $M$ skips the current cycle. If the number of continuously skipped cycles exceeds the limit $K$, the processor $M$ shuts the system down. When a transient fault occurs in a processor, rebooting the processor repairs the fault. Lastly, if the main processor fails, the system is automatically shut down. The mean times to failure for the sensors, actuators, I/O processors, main processor, and the mean times for the delays are given in Table \ref{tab:observedvariables}, in which $1$ time unit is $30$ seconds. A cycle lasts $2$ time units, that is 1 minute.

As described above, the system is modeled as a SHLPN. In the net, the places for each sensor group and each actuator group have $4$ and $3$ different markings, respectively. The places for I/O processors have $3$ different marking, and the place for main processor have $2$ different marking. Therefore, the underlying CTMCs for the net has $\sim 2^{155}$ states comparing to the model in \cite{knp07} with $\sim 2^{10}$ states.

\subsection{Analysis Results}
The set of observed variables, the temporal resolution and their meaning are given in Table \ref{tab:observedvariables}. We define four types of failures: $\mathtt{failure_1}$ is the failure of the sensors, $\mathtt{failure_2}$ is the failure of the actuators, $\mathtt{failure_3}$ is the failure of the I/O processors and $\mathtt{failure_4}$ is the failure of the main processor. For example, $\begin{tt}failure_1\end{tt}$ is defined as $(\begin{tt}number\_sensors\end{tt} < 37) \wedge (\begin{tt}proci\_status\end{tt} = 2)$. It specifies that the number of working sensor groups has decreased below $37$ and the input processor is functional, so that it can report the failure to the main processor. We define $\begin{tt}failure_2\end{tt}$, $\begin{tt}failure_3\end{tt}$, and $\begin{tt}failure_4\end{tt}$ in a similar way. In our analysis which is based on the one in \cite{knp07} with $K = 4$, we used the Monte Carlo algorithm with $3000$ simulations.
\begin{table}[ht]
\centering
\footnotesize
\ra{1.0}
\begin{tabular}{@{}lclclcl@{}}
\toprule
Variable name & \phantom{abc} & Meaning & \phantom{abc} & Component & \phantom{abc} & Mean time\\
\midrule
$\begin{tt}number\_sensors\end{tt}$ & \phantom{abc} & \textit{Working sensor groups} & \phantom{abc} & $\begin{tt}Sensor\end{tt}$ & \phantom{abc} & \textit{1 month}\\
$\begin{tt}number\_actuators\end{tt}$ & \phantom{abc} &	\textit{Working actuator groups} & \phantom{abc} & $\begin{tt}Actuator\end{tt}$ & \phantom{abc} & \textit{2 months}\\
$\begin{tt}proci\_status\end{tt}$ & \phantom{abc} & \textit{Input processor's state} & \phantom{abc} & $\begin{tt}Transient\end{tt}$ & \phantom{abc} & \textit{1 day}\\
$\begin{tt}proco\_status\end{tt}$ & \phantom{abc} & \textit{Output processor's state} & \phantom{abc} & $\begin{tt}Processor\end{tt}$ & \phantom{abc} & \textit{1 year}\\
$\begin{tt}procm\_status\end{tt}$ & \phantom{abc} & \textit{Main processor's state} & \phantom{abc} & $\begin{tt}Timer cycle\end{tt}$ & \phantom{abc} & \textit{1 minute}\\
$\begin{tt}timeout\_counts\end{tt}$ & \phantom{abc} & \textit{Number of skipped cycles} & \phantom{abc} & $\begin{tt}Reboot\end{tt}$ & \phantom{abc} & \textit{30 seconds}\\
$\begin{tt}reward\_up\end{tt}$ & \phantom{abc} & \textit{Time in ``up'' state} & & & &\\
$\begin{tt}reward\_danger\end{tt}$ & \phantom{abc} & \textit{Time in ``danger'' state} & & & &\\
$\begin{tt}reward\_shutdown\end{tt}$ & \phantom{abc} & \textit{Time in ``shutdown'' state}& & & &\\
\midrule
Temporal resolution & \phantom{abc} & Meaning &&&&\\
\midrule
$\begin{tt}tick\_notified\end{tt}$ & \phantom{abc} & \multicolumn{5}{l}{\textit{Observed variables are evaluated every one time unit}}\\
\bottomrule
\end{tabular}
\caption{Observed variables and temporal resolution}
\label{tab:observedvariables}
\end{table}

\begin{figure}[ht]
\centering
\includegraphics[width=0.8\textwidth]{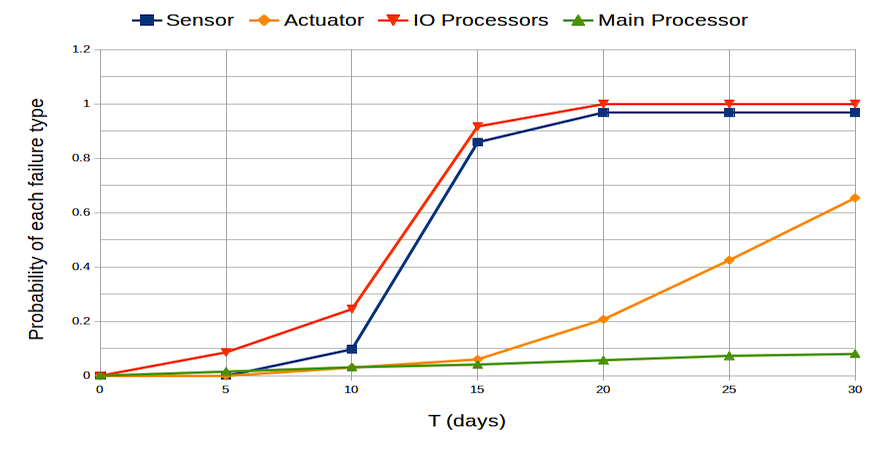}
\caption{The probability that each of the 4 failure types in the first $T$ time of operation}
\label{fig:failure}
\end{figure}

First, we study the probability that each of the four types of failure eventually occurs during the first $T$ units of time using the formula $F_{\leq T} \; (\begin{tt}failure_i\end{tt})$. Fig.~\ref{fig:failure} plots these probabilities for $T$ varying from $5$ to $30$ days of operation. We observe that the probabilities that the sensors and I/O processors eventually fail are higher than the others.

\begin{figure}[ht]
\centering
\includegraphics[width=0.8\textwidth]{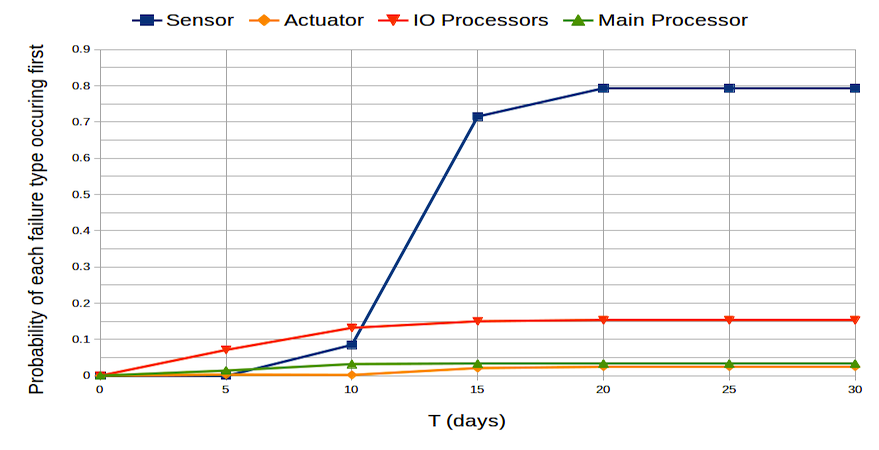}
\caption{The probability that each of the 4 failure types is the cause of system shutdown in the first $T$ time of operation}
\label{fig:failurefirst}
\end{figure}

For the second part of our analysis, we try to determine which kind of component is more likely to cause the failure of the system. In that frame, it is necessary to determine the probability that a failure related to a given component occurs before any other failures. The atomic proposition $\begin{tt}shutdown\end{tt}$, defined by $\bigvee_{i=1}^4\begin{tt}failure_i\end{tt}$ indicates that the system has shut down because one of the failures has occurred. The formula $\neg \begin{tt}shutdown\end{tt} \; U_{\leq T} \; \begin{tt}failure_i\end{tt}$ is true if the failure $i$ occurs within $T$ time units and no other failures have occurred before the failure $i$ occurs, that is if failure $i$ is the cause of the shutdown. Fig.~\ref{fig:failurefirst} shows the probability that each kind of failure occurs first over a period of $30$ days of operation. It is obvious that the sensors are likelier to cause a system shutdown. At $T=20$ days, it seems that we reached a stationary distribution indicating for each kind of component the probability that it is responsible for the failure of the system.

\begin{figure}[ht]
\centering
\includegraphics[width=0.8\textwidth]{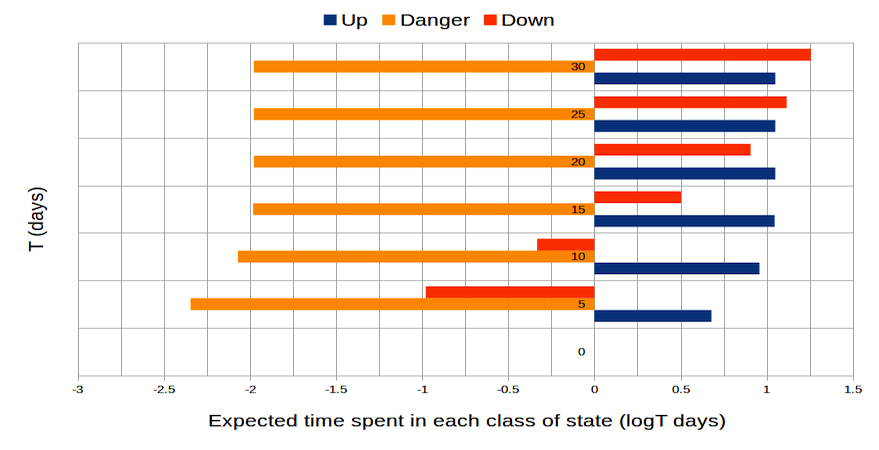}
\caption{The expected amount of time spent in each of the states: ``up'', ``danger'' and ``shutdown''}
\label{fig:timespent}
\end{figure}

For the third part of our analysis, we divide the states of system into three classes: ``up'', where every component is functional, ``danger'', where a failure has occurred but the system has not yet shut down (e.g., the I/O processors have just had a transient failure but they have rebooted in time), 
and ``shutdown'', where the system has shut down \cite{knp07}. We aim to compute the expected time spent in each class of states by the system over a period of $T$ time units. 
To this end, we add in the model, for each class of state $\mathtt{c}$,  a variable $\begin{tt}reward\_c\end{tt}$ that measures the time spent in the class $\mathtt{c}$. 
The formula $X_{\leq T} \; \begin{tt}reward\_c\end{tt}$ returns the mean value of  $\mathtt{reward\_c}$ after $T$ time units of execution over the 3000 traces considered. The results are plotted in Fig.~\ref{fig:timespent}.

\begin{figure}[ht]
\centering
\includegraphics[width=0.8\textwidth]{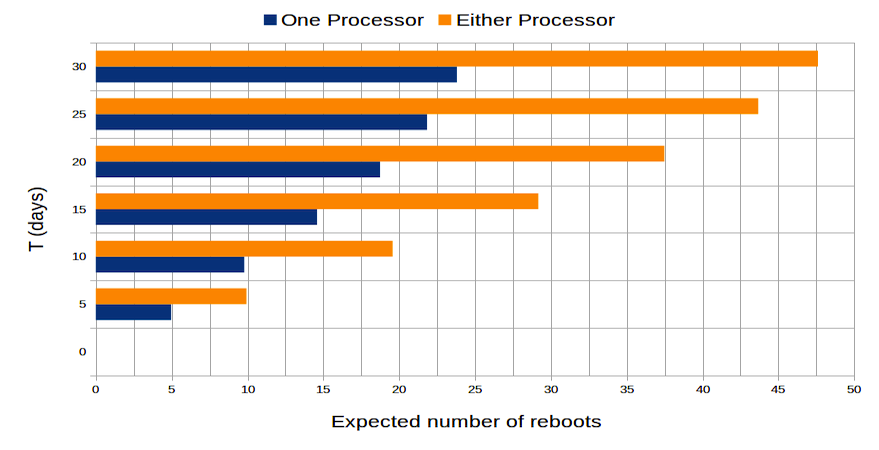}
\caption{Expected number of reboots that occur in the first $T$ time of operation}
\label{fig:numberofreboots}
\end{figure}

\begin{figure}[ht]
\centering
\includegraphics[width=0.8\textwidth]{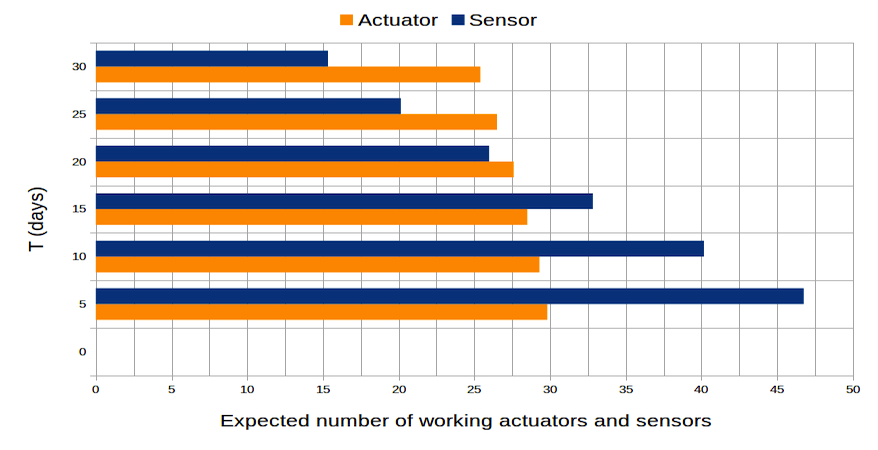}
\caption{Expected number of functional sensor and actuator groups in the first $T$ time of operation}
\label{fig:numberofworkinggroups}
\end{figure}

Finally, we approximate the number of input, output processor reboots which occur and the number sensor groups, actuator groups that are functional over time by computing the expected values of random variables that count the number of reboots, functional sensor and actuator groups. The results are plotted in Fig. \ref{fig:numberofreboots} and Fig. \ref{fig:numberofworkinggroups}. It is obvious that the number of reboots of both processors doubles the number of reboots of each processor since they have the same models.

\section{Related Work and Conclusion}
\label{sec:conclusion}
Some work has been carried out for dependability analysis with PMC, for example, the dependability analysis of control system with PRISM~\cite{knp07}. PRISM supports construction and analysis of Markov chains. 
For example, the exact probabilities in our case study can be computed  by PRISM for the small system with one sensor group and one actuator group. However, the main drawback of this approach is that when it deals with real-world large size systems which make the PMC technique is unfeasible. That means the state explosion likely occurs, even with some abstraction, i.e., symbolic model checking with \textit{Ordered Binary Decision Diagrams} (OBDDs), is applied.

Tabakov et al.~\cite{tva10} proposed a framework for monitoring temporal SystemC properties. This framework allows users express the verifying properties by fully exposing the semantics of the simulator as well as the user-code. They extend LTL by providing some extra primitives for stating the atomic propositions and let users define a much finer temporal resolution. Their implementation consists of a modified simulation kernel, and a tool to automatically generate the \textit{monitors} and aspect advices for applying \textit{Aspect Oriented Programming} (AOP) \cite{gss01} to instrument SystemC programs automatically.

This paper presents the first attempt to analyze the dependability of computer-based control systems using statistical model checking, in which the dependability of the systems is modeled by a SystemC-realization of stochastic high-level Petri nets. In comparison to the probabilistic model checking, our approach allows users to handle large industrial systems as well as to expose a rich set of user-code primitives by automatically instrumenting the SystemC code with AspectC$^{++}$.

Currently, we consider an external library as a ``black box'', meaning that we do not consider the states of external libraries. Thus, arguments passed to a function in an external library cannot be monitored. For future work, we would like to allow users to monitor the states of the external libraries with the future version of AspectC$^{++}$. We also plan to apply statistical model checking to verify temporal properties of SystemC-AMS (Analog/Mixed-Signal) and make the improved version of the current naive inference algorithm implementation based on the RETE algorithm in~\cite{for82}.

\bibliographystyle{abbrv}
\bibliography{RR-8762}

\end{document}